# A Multiwavelength Strategy for Identifying Celestial γ-ray Sources

Patrizia A. Caraveo

*Istituto di Fisica Cosmica "G. Occhialini"*
*Via Bassini, 15 –20133 Milano, Italy*
*pat@ifctr.mi.cnr.it*

**Abstract.** The vast majority of the high-energy γ-ray sources discovered by EGRET are still unidentified. Percentages range from 50% at high galactic latitudes, where blazars are responsible for almost all identified sources, to more than 90% near the galactic plane, where isolated neutron stars appear to be the only certified class of sources of high energy γ-rays. In spite of all the efforts devoted to the identification problem, the only success story, so far, appears to be the chase for Geminga, where X-rays led the way to eventual optical identification. Similar searches are now starting to produce encouraging results, although none has reached, as yet, a certified identification.

## HISTORICAL OVERVIEW

Unidentified sources are as old as γ-ray astronomy itself. As soon as the NASA SAS-2 satellite was able to discriminate point-like sources from the underlying diffuse emission, an unidentified source appeared in the γ-ray sky. The very first images of the galactic anticentre unveiled γ 195+5, later to become Geminga. When, in 1973, the SAS-2 mission ended prematurely, γ-ray sources, encompassing the Crab and Vela pulsar together with the unidentified one, were a reality (Fichtel et al, 1975).

COS-B continued on the same track. Its longer active life span allowed for a significant increase in the number of photons collected, yielding a grand total of 25 sources, almost a ten-fold increase with respect to SAS-2. The second COS-B catalogue is the legacy to high energy astronomy of a successful mission. It recognized, for the first time, the paramount importance of unidentified objects which, at the time, appeared to be responsible for 21 of the 25 detections. Of course, the degree-size positional uncertainty of each source, prevented straightforward identifications. Only objects with unambiguous timing signature (such as pulsars) could be reliably identified. Indeed, apart from the Crab and Vela pulsars, COS-B was able to pinpoint two prominent objects in its error boxes, namely the first extragalactic source, 3C273, and the molecular cloud ρ-Oph. (see Bignami & Hermsen, 1983, for a review). In spite of the COS-B uneven sky coverage, well visible in Figure 1, the concentration of sources along the galactic plane appeared to be real, pointing towards a galactic population with a rather small scale height and an average distance between 2 and 7 kpc, implying an average luminosity in the range $(0.4-5)10^{36}$ erg/sec (Swanenburg et al, 1981).

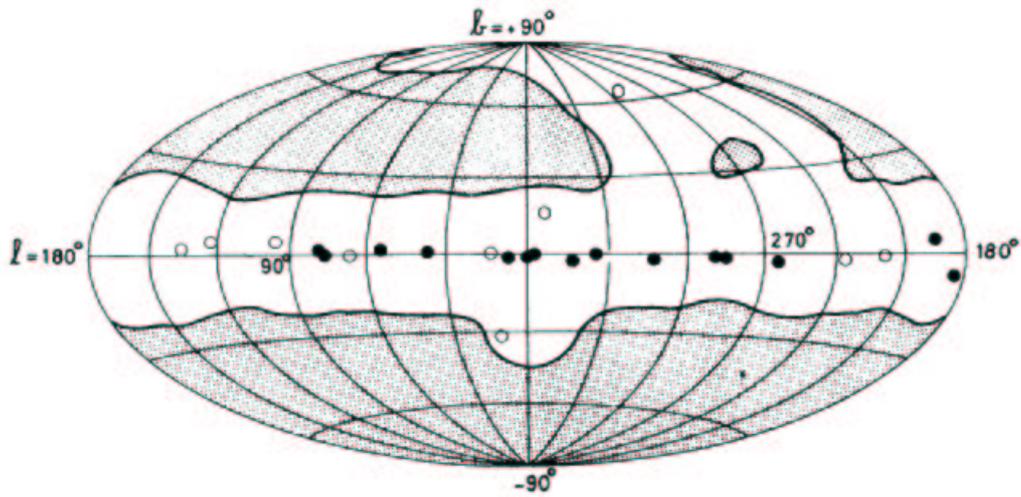

**FIGURE 1.** Second (and last) COS-B Catalogue of γ-ray sources (Swanenburg et al. 1981). Open circles denote faint sources, filled circles brighter ones. The dividing line is at $1.3 \cdot 10^{-6}$ ph/cm$^2$sec. The shaded area has not been searched for sources since the coverage was not adequate.

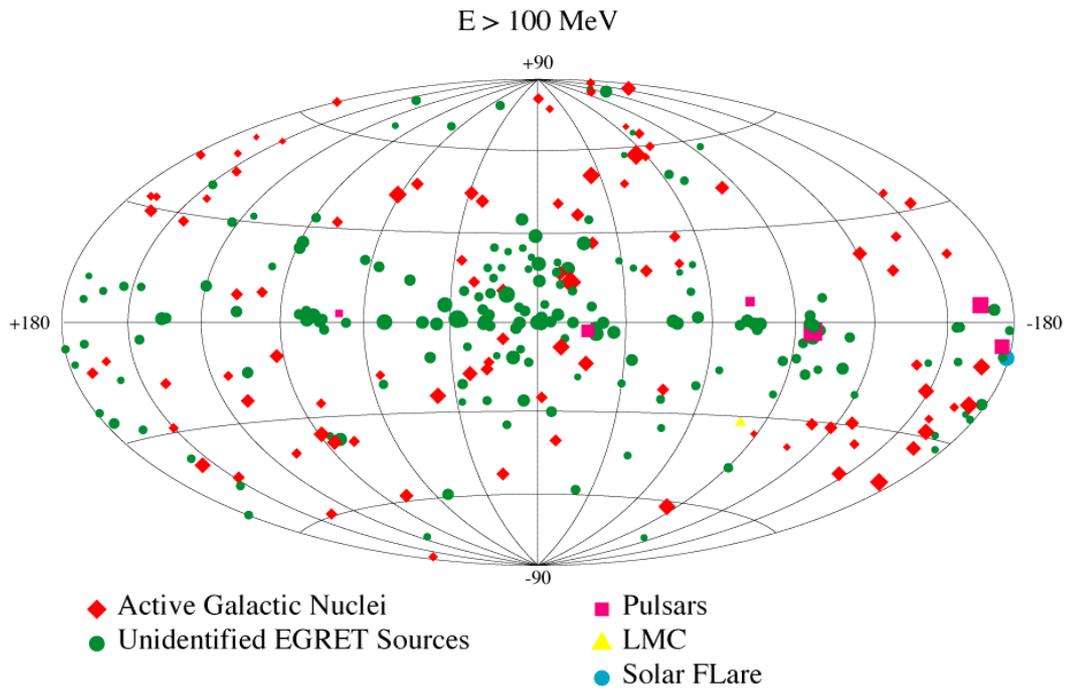

**FIGURE 2.** Third EGRET catalogue (Hartman et al., 1999). The size of the symbols, the meaning of which is spelled out in the legenda, is proportional to the measured source flux.

EGRET, onboard the Compton Gamma ray Observatory, provided another ten-fold increase in the number of sources (see Figure 2). Owing to a much bigger sensitive area, coupled with a better angular resolution, EGRET was able to lower significantly the COS-B detection limit. Moreover, by covering the entire sky, EGRET unveiled a new class of powerful γ-ray emitters: the Blazars . At variance with low latitude galactic sources, Blazars are best seen in high latitude, less crowded fields. Often, their γ-ray flux shows variability correlated to the radio and optical ones, making the identification job much more easy. Unfortunately, the identifications of the low latitude sources continue to be hampered by the error box dimensions. Although smaller than the COS-B ones, the EGRET error boxes are still too big to allow for direct identifications. Statistical studies on the low latitude sources, performed by Mukherjee et al (1995), confirmed the findings of Swanenburg et al, (1981). The unidentified low latitude EGRET sources are at distances between 1.2 and 6 kpc and their luminosities are in the range (0.7- 16.7) $10^{35}$ erg/sec.

It is worth mentioning that EGRET discovered short-term variability from GRO J1838-04 (Tavani et al ,1997), later listed in the EGRET catalogue as 3CGJ1837-0423. However, the lack of any real time notification of such a variability hampered the search of an equally varying counterpart, thus missing a precious chance for source identification.

Table 1 summarizes the source growth-rate from COS-B to EGRET. The low latitude (b<10°), presumably galactic, sources went from 22 to 80, while the high latitude (b>10°), presumably extragalactic, ones jumped from 3 to 181, with Blazars accounting for roughly half of them. If we forget this classical (and somewhat arbitrary) separation and look at the distribution of all the EGRET sources, we realize that there is an excess of sources at middle latitudes. Gehrels et al (2000) have proposed this to be a new classes of γ-ray sources, possibly linked to the Gould's Belt structure (see also Grenier, 2000). On average, middle latitude sources are fainter than the low latitude ones, closely aligned with the galactic plane, and their spectral shapes tend to be softer. Moreover, middle latitude sources should be closer to us than low latitude ones, thus their average luminosity should be lower.

Among the 80 low latitude sources, Table1 shows how pulsars numbered from 2 to 6, with three more possible entries (Thompson, 2001). However important, such an increase confirms pulsars (be they radio loud or radio quiet) as an established class of gamma-ray emitters, although, probably not the only one. Indeed, the comparison between the measured pulsar luminosities (e.g. Thompson et al., 1999) and the average values inferred from the statistical study shows clearly that only very young and energetic pulsars could fulfil the luminosity requirements of unidentified sources. This limits quite severely the total contribution of classical pulsars to the low latitude, galactic γ-ray source population

What about the remaining unidentified sources? Surprisingly enough, EGRET has done very little to clarify the nature of the galactic γ-ray emitters. While their number increased fourfold, successful identifications remain at a meagre <10% level. They encompass, anyway, only isolated neutron stars.

**TABLE 1. From COS-B to EGRET.**

| Low Latitude (b<10°) | COS-B | EGRET |
|---|---|---|
| Total | 22 | 80 (- 1 solar flare) |
| Identified | **pulsars** | **pulsars** |
| | Crab, Vela | Crab, Vela, PSR 1706-44, PSR1951+32, PSR1055-52 |
| | | Geminga (**radio quiet**) |
| **% unidentified** | **90%** | **90%** |

| High Latitude (b>10°) | COS-B | EGRET |
|---|---|---|
| Total | 3 | 181 |
| Identified | 3C273, ρ Oph | 66 *variable* **AGNs** |
| | | 28 probable **AGNs** |
| | | LMC, CenA |
| **% unidentified** | - | **50%** |

# COUNTERPART SEARCHES

## More of the same: Pulsars

In view of the overwhelming presence of pulsars among identified galactic γ-ray sources, it is natural to explore the possibility that at least a fraction of the remaining low latitude sources belong to the same class of compact objects.

Indeed, the search for new pulsars started as soon as COS-B discovered a population of unidentified sources, but no pulsars were unveiled. New searches were triggered by the EGRET findings, but the lack of results, experienced at the time of COS-B, appears to be unchanged. Dedicated radio searches (Nice and Sayer, 1997), aimed precisely at the search for radio pulsars inside the error boxes of 10 of the brightest EGRET sources, yielded null results. The implication is that straightforward radio pulsar identification is not the only possible solution to the enigma of the unidentified high-energy γ-ray sources. This is further strengthened by the work of Nel et al. (1996) who investigated 350 known pulsars. They found few positional coincidences but no significant γ-ray timing signature for any of the pulsars in the survey.

However, more powerful surveys, now underway, may be changing this negative trend. Kaspi et al.(2000) find evidence of an association between the 20,000 y old PSR B1046-58, one of the candidates listed by Thompson (2001), and 3EG J1048-5840. D'Amico et al. (2001), using the new Parkes data, discover two young, promising radio pulsars inside the error boxes of 3EG J1420-6038 and 3EG J1837-0606. In view of the time noise, usually present in young pulsars, it will be difficult to search for their time signature in the EGRET data. They will be certainly studied by the future gamma ray missions, as was the case for the unidentified COS-B source 2CG 342-02, identified by EGRET with the newly discovered PSR 1706-49.

## An Elusive Template

Apart from classical radio pulsars, γ-ray astronomy does offer a remarkable example of an isolated neutron star (INS) which behaves as a pulsar as far as X-and-γ-astronomy are concerned but has little, if at all, radio emission. As an established representative of the non-radio-loud INSs (see Caraveo, Bignami and Trümper, 1996 for a review), Geminga offers an elusive template behaviour: prominent in high energy γ-rays, easily detectable in X-rays and downright faint in optical, with sporadic or no radio emission (see Bignami and Caraveo, 1996 for a review of the source multiwavelength phenomenology).

Although the energetic of Geminga ($L_\gamma = 3.3 \; 10^{34}$ erg/sec) is not adequate to account for the very low latitude (i.e. more distant) EGRET sources, it could satisfy the fainter middle latitude sources, presumably belonging to a more local galactic population (Gehrels et al. 2000). Their gamma yield is certainly compatible with the rotational energy loss of a middle-aged neutron star, like Geminga.

Thus, in spite of being the only confirmed identification, Geminga does not provide a viable template for the entire family of unidentified γ-ray sources. Some other object or class of objects is needed. However, the strategy devised for the chase of Geminga seems to still be the best one to bridge the positional accuracy gap intrinsic to γ-ray astronomy.

## Ongoing Multiwavelength Efforts

The γ-to-X-to-optical multiwavelength approach has by now been applied to a number of COS-B and EGRET sources.

The Einstein coverage of the error box of 2CG 135+01 yielded the discovery of the X-ray emission of the periodically variable radio source GT61.303 (Bignami et al, 1981). In spite of ad hoc searches for γ-ray variability correlated with the radio one (e.g. Kniffen et al, 1997; Tavani et al., 1998), no conclusive proof has been brought forward for the identification of 2CG135+1, now 3EG 0241+6103, with this peculiar binary system. The association remains, however, tantalizing.

For 3EG J0634+0521, Kaaret et al. (1999) have suggested the identification with SAX J0635+0533, a binary system containing a compact object with a Be star companion. Recently, such a proposed identification has been strengthened by the discovery of a 34 msec pulsation (Kareet, Cusumano and Sacco, 2000), coupled with a high period derivative, pointing towards a young, energetic pulsar in a binary system. This is an absolute first in the panorama of known binary systems, making SAX J0635+0533 an interesting system "per se". For both 3EG J0241+6103 and 3EG J0634+0521, the physics behind the γ-ray production would be particle acceleration at the shock created by the pulsar interaction with the thick Be-star wind during periastron passage.

Roberts, Romani and Kawai (2001) list a number of pulsar nebulae which could be associated, mostly on positional grounds, with their specially selected GeV sources. Of special interest is the case of the Kookaburra Nebula, in the error box of 3EG J 1420-6038, where is located also one of the new pulsars of D'Amico et al. (2001). Oka et al (1999) invoke the interaction of a pulsar nebula with a dark cloud to account for 3EG J 1809-2328.

A newly discovered, energetic, young, isolated pulsar, showing 51.6 msec X-ray and radio pulsations, has been proposed by Halpern et al (2001a) as the counterpart of 3EG J2227+6122. As we have already remarked in the case of the young pulsars discovered by D'Amico et al (2001), only future missions will be able to confirm such a promising identification by detecting the pulsar time signature in γ-rays.

Radio quiet INS identifications have been proposed for 3EG J1835+5918, the brightest unidentified γ-ray source, (Mirabal and Halpern 2001; Reimer et al., 2001), 3EG J0010+7309 (Brazier et al, 1998) and 3EG J2020+4026 (Brazier et al., 1996) . It is interesting to note that both 3EGJ 1835+5918 and 3EG J0010+7309 are middle latitude sources, thus their energetic requirements are easily compatible with a Geminga-like identification.

3EG J2016+3657, on the other hand, has been identified with a Blazar behind the Galactic plane (Mukherjee et al, 2000; Halpern et al, 2001b). This extragalactic "contamination", deep in the galactic plane, should not come as a surprise. The isotropic distribution of Blazars, coupled with the negligible absorption suffered by γ-ray photons through the galactic plane, should result in quite a few galactic Blazars.

Thus, years of multiwavelength world-wide efforts have yielded the following counterparts:

**4 energetic young radio pulsars**, one of which has been discovered in X-rays. So far, only PSR B1046-58 is considered a candidate EGRET pulsar;
**3 Geminga-like**, radio quiet INSs;
**2 peculiar binary systems** (one with a young pulsar);
**1 galactic Blazar;**
few pulsar nebulae.

All in all, about a dozen EGRET sources have a tentative, more-or-less likely, identification. If compared to the scores of sources awaiting an identification, the number is small, however, the panorama is a rapidly evolving one. Two years ago, a similar review would have resulted in not more than 4 tentative identifications. Indeed, the most recent (and sometimes most promising) entries in our summary list are the outcome of the searches prompted by the second EGRET catalogue (Thompson et al. 1995, 1996). This gives us an idea of the time needed to bring to completion a thorough search for counterparts requiring one, or more, cycles of X-rays observations followed by one, or more, cycles of optical (and radio) ones.

The multiwavelength approach, although promising, is a time consuming exercise.

## THE NEW MILLENIUM

New instruments, now operational both in the optical and X-ray domains, promise to speed up significantly the X and optical coverage of γ-ray error boxes.

On the X-ray side, two great observatories such as Chandra and Newton-XMM can cover with few, relatively short, pointings each EGRET error box, pushing the source detection limit to unprecedented levels.

It is easy to predict that dozens of serendipitous sources (mainly stars and AGNs) will be detected in each EGRET error box by these powerful X-ray telescopes.

Thus, it will be critical to plan for a "massive" approach to the optical identification work, which is bound to become the bottleneck of this multiwavelength chain.

Optical wide-field-imagers, with typical field-of-views of a fraction of square degree, offer just such a new perspective since they can speed-up considerably the tedious X-to-optical comparison work, aimed at discarding candidates with obvious identifications.  Moreover, taking advantage of field of views comparable to those of the X-ray telescope,  the optical work can proceed independently from the X-ray one, bypassing the need to wait for the results of the X-ray observations to plan the optical follow-up exposures.

In spite of the new instruments, selecting promising targets will remain a difficult matter. So far, γ-ray astronomy has not been able to classify its sources in families characterized by different templates at various wavelengths. Source classification should proceed together with the identification work and should be based on a trial-and-error approach.

## Our Program

For our  Newton-XMM exploratory program, we shall apply the Geminga strategy to two middle latitude EGRET sources. First of all, one has to single out potential neutron star candidates taking advantage of the high $F_x/F_v$ of these objects. All efforts must then converge towards the identification of the neutron star candidate.

Our two EGRET sources, namely 3EG J1249-8330 and 3EG J0616-3310, have been selected on the basis of their positional accuracy, spectral shape, galactic location and lack of candidate extragalactic counterparts. Each source error box, a circle of 30' radius,  will be covered by EPIC with four 10,000 sec exposure pointings, yielding a homogeneous coverage of about 1 sq deg. On the basis of similar pointings, we expect to have 50-100 sources, as faint as $10^{-14}$ erg/cm$^2$ sec,  in each EGRET error box.

The X-ray coverage will be complemented by the optical one, done in three colours by the European Southern Observatory WFI (Wide Field Imager), operating at the ESO 2.2m telescope in La Silla, Chile. Its 8 CCD detectors cover  a 30' x 30' f.o.v., a value directly comparable to the EPIC one.  In order to be able to discard non-neutron-star optical IDs, we ought to reach a limiting magnitude of $m_v$ 25, a value typical for  a 1 h dithered exposure. Thus, in order to do a multicolour optical coverage of a 10,000 sec EPIC exposure, we will need about the same observing time with a 2 m class optical telescope. Cross correlation of the optical and X-ray images will yield the $F_x/F_v$ parameter for the X-ray sources discovered by EPIC. Different colours will provide a further handle to solve ambiguous cases, where more than one optical entry will be compatible with the X-ray position

Of course, should new templates arise, they could be immediately implemented taking advantage of our unbiased X and optical coverage.

# CONCLUSIONS

Irrespective of the nature(s) of unidentified γ-ray sources, it is clear that we are still far from a general solution, if any. A lot of observing time, both in X-rays and in the optical, should be devoted to these mysterious objects, in order to identify as many as possible of them before future high energy missions will start their active life. With some a-priori knowledge of the nature of the sources they are going to observe, missions such as Agile or Glast could optimize their observing strategies